# Giant nonlinear photocurrent in $\mathcal{PT}$-symmetric magnetic topological quantum materials

Hua Wang[1] and Xiaofeng Qian[1]*

**Affiliations:**
[1]Department of Materials Science and Engineering, Texas A&M University, College Station, TX 77843, USA

*Correspondence to: feng@tamu.edu

**Abstract:**

Recently, nonlinear photocurrent such as normal shift photocurrent (NSC) and normal injection photocurrent (NIC) in time-reversal invariant noncentrosymmetric systems have attracted substantial interest. Here we propose two new types of second-order nonlinear direct photocurrents, magnetic shift photocurrent (MSC) and magnetic injection photocurrent (MIC), as the counterparts of NSC and NIC in time-reversal symmetry and inversion symmetry broken system. We show that MSC is mainly governed by shift vector and interband Berry curvature, and MIC is dominated by absorption strength and asymmetry of the group velocity difference at time-reversed $\pm \boldsymbol{k}$ points. MSC and MIC can be induced by circularly and linearly polarized light in $\mathcal{PT}$-symmetric system with $\mathcal{P}$ and $\mathcal{T}$ being individually broken, respectively. Taking $\mathcal{PT}$-symmetric magnetic topological quantum material bilayer antiferromagnetic (AFM) $MnBi_2Te_4$ as an example, we predict the presence of large MIC in the terahertz frequency regime which can be switched between two AFM states with time-reversed spin orderings upon magnetic transition. In addition, while NSC vanishes in $\mathcal{PT}$-symmetric system, external electric field breaks $\mathcal{PT}$ symmetry and enables large NSC response in bilayer AFM $MnBi_2Te_4$ which can be switched by external electric field. More importantly, due to the magnetic point group symmetry of bilayer AFM $MnBi_2Te_4$, MIC and NSC are perpendicular to each other upon linearly $x/y$-polarized light, hence can be distinguished. Furthermore, both MIC and NSC are highly tunable under varying electric field due to the field-induced large Rashba and Zeeman splitting, resulting in giant nonlinear photocurrent response down to a few THz regime, suggesting bilayer AFM-$z$ $MnBi_2Te_4$ as a tunable platform with rich THz and magneto-optoelectronic applications. The present work reveals that nonlinear photocurrent responses governed by NSC, NIC, MSC, and MIC provide a powerful tool for deciphering magnetic structures and interactions which could be particularly fruitful for probing and understanding magnetic topological quantum materials.

**Main**

Magnetic topological quantum materials have attracted a lot of interest as the interplay between magnetic ordering and topology may enable exotic topological quantum states. For example, antiferromagnetic topological insulator [1] can host topological axion states with nonzero quantized Chern-Simons magnetoelectric coupling [2] which was recently predicted by first-principles theory in the class of layered $MnBi_2Te_4$ materials [3-5]. The nontrivial topological surface states have been experimentally verified in $MnBi_2Te_4$ [6-8]. While various experiments have been carried out to study their electronic structure and transport properties, optical probes, especially nonlinear optical spectroscopy, may provide a rich set of alternative tools to probe their topological nature and understand the inherent electronic structure in these exotic quantum materials.

Nonlinear optical responses play a key role in understanding the symmetry, ordering, and topology. Materials with strong nonlinear responses are particularly valuable for ultrafast nonlinear optics [9], efficient generation of entangled photon pairs [10,11], and phase-matching free nonlinear optics [12,13] using 2D materials. Recently, nonlinear photocurrent responses including normal shift current (NSC) [14-17], normal injection current (NIC) [17-19], and Berry curvature dipole induced current [20-22] have been observed in time-reversal invariant noncentrosymmetric systems. In particular, Berry curvature dipole induced photocurrent is closely coupled with ferroelectric orders, offering a unique approach to detect low-energy ferroelectric transition. This was recently theoretically proposed and experimentally verified in ferroelectric few-layer $WTe_2$ semimetal, opening avenues to the development of nonlinear quantum memory [22,23]. Compared to few-layer $WTe_2$, layered $MnBi_2Te_4$ is magnetic with tunable nontrivial topology, which may lead to magnetically and electrically controlled large nonlinear photocurrent responses. The effect of magnetic ordering and intrinsic topology on nonlinear responses in $MnBi_2Te_4$, however, has been largely underexplored.

Here we propose two new types of second-order nonlinear current responses, namely *magnetic injection current* (MIC) and *magnetic shift current* (MSC), which are the counter part of the well-known NIC and NSC in time-reversal invariant noncentrosymmetric system. Microscopic theory and group theoretical analysis of nonlinear photocurrent, especially the two new MIC and MSC responses, are provided. Both MIC and MSC can be induced in $\mathcal{PT}$-symmetric systems with $\mathcal{P}$ and $\mathcal{T}$ being individually broken. We take magnetic topological quantum material bilayer antiferromagnetic (AFM) $MnBi_2Te_4$ as an example of $\mathcal{PT}$-symmetric system, and predict that MIC can be switched in time-reversal breaking $MnBi_2Te_4$ upon magnetic transition, namely magnetically switchable nonlinear photocurrent. In addition, while $\mathcal{PT}$ symmetry forbids the NSC response, vertical electric field can easily break the $\mathcal{PT}$ symmetry and enable large NSC in bilayer AFM $MnBi_2Te_4$. Both MIC and NSC are highly tunable under varying electric field due to the field-induced large Rashba and Zeeman splitting. Electric field induced bandgap reduction not only leads to topological phase transition of bilayer AFM $MnBi_2Te_4$ from zero-plateau quantum anomalous Hall insulator into high Chern number quantum anomalous Hall insulator ($C_n = 3$), but also results in giant nonlinear photocurrent response redshifted down to a few THz regime, suggesting $MnBi_2Te_4$ may serve as a promising platform for terahertz sensing and magneto-optoelectronic applications. More generally, the rich nonlinear photocurrent responses in time-reversal and inversion symmetry broken systems are inherently coupled with their crystal structure and magnetic symmetry, offering a promising route for probing and understanding the intrinsic electronic structure and potentially topological physics in magnetic quantum materials.

**Microscopic theory of second-order nonlinear photocurrent**

The general second-order direct injection current (IC) and shift current (SC) under monochromatic electric field $E^b(t) = E^b(\omega_\beta)e^{-i\omega_\beta}$ with $\omega_\beta = \pm\omega$ in the clean limit are given by [14,17,24]

$$\frac{dJ_{\text{IC}}^a}{dt} = -\frac{\pi e^3}{\hbar^2} \int [d\mathbf{k}] \sum_{mn\sigma} f_{mn} \Delta_{mn} r_{nm}^b r_{mn}^c \, \delta(\omega_{nm} - \omega_\beta) E^b(\omega_\beta) E^c(-\omega_\beta),$$

$$J_{\text{SC}}^a = -\frac{i\pi e^3}{2\hbar^2} \int [d\mathbf{k}] \sum_{mn\sigma} f_{nm} \left( r_{mn}^b r_{nm;k^a}^c - r_{nm}^c r_{mn;k^a}^b \right) \delta(\omega_{nm} - \omega_\beta) E^b(\omega_\beta) E^c(-\omega_\beta),$$

where $a, b, c$ are Cartesian indices, $r_{mn;k^a}^b = \frac{\partial r_{mn}^b}{\partial k^a} - i r_{mn}^b (\mathcal{A}_m^a - \mathcal{A}_n^a)$ is the gauge covariant derivative. $\mathbf{r}_{nm} = i\langle n|\partial_\mathbf{k}|m\rangle$ and $\mathcal{A}_n = i\langle n|\partial_\mathbf{k}|n\rangle$ are interband and intraband Berry connection, respectively. $f_m$ is the Fermi-Dirac distribution with $f_{nm} \equiv f_n - f_m$, and $\hbar\Delta_{mn}^a \equiv v_{mm}^a - v_{nn}^a$ is the group velocity difference of band $m$ and $n$. $[d\mathbf{k}] \equiv \frac{d\mathbf{k}}{(2\pi)^d}$ for $d$ dimension. Here $e = -|e|$ carries a negative sign due to the negative charge of electron. We separate the product of the electric field amplitude into the symmetric real and antisymmetric imaginary parts,

$$E^b(\omega)E^c(-\omega) = E^b(\omega)E^{c*}(\omega) = \text{Re}(E^b E^{c*}) + i\text{Im}(E^b E^{c*}),$$

$$E^b(-\omega)E^c(\omega) = E^{b*}(\omega)E^c(\omega) = \text{Re}(E^b E^{c*}) - i\text{Im}(E^b E^{c*}).$$

For linearly polarized light, $\mathbf{E}(\omega)$ is real, while for left/right-circularly polarized light $\mathbf{E}(\omega)$ is complex with $\text{Im } E^b(\omega)E^c(-\omega) = -\text{Im } E^b(-\omega)E^c(\omega) \neq 0$. After symmetrization with $\pm\omega$, we have

$$\frac{dJ_{\text{IC}}^a}{dt} = -\frac{\pi e^3}{\hbar^2} \int [d\mathbf{k}] \sum_{mn\sigma} f_{mn} \Delta_{mn} \{r_{nm}^b, r_{mn}^c\} \delta(\omega_{nm} - \omega) \, \text{Re}(E^b E^{c*}) -$$
$$i\frac{\pi e^3}{\hbar^2} \int [d\mathbf{k}] \sum_{mn\sigma} f_{mn} \Delta_{mn}^a [r_{nm}^b, r_{mn}^c] \delta(\omega_{nm} - \omega) \, \text{Im}(E^b E^{c*}),$$

where $\{r_{nm}^b, r_{mn}^c\} \equiv r_{nm}^b r_{mn}^c + r_{nm}^c r_{mn}^b$, and $[r_{nm}^b, r_{mn}^c] \equiv r_{nm}^b r_{mn}^c - r_{nm}^c r_{mn}^b$. We then arrive at

$$\frac{dJ_{\text{IC}}^a}{dt} = 2\eta_{\text{MIC}}^{abc} \text{Re}(E^b E^{c*}) + 2i\eta_{\text{NIC}}^{abc} \text{Im}(E^b E^{c*}),$$

where

$$\eta_{\text{MIC}}^{abc} = -\frac{\pi e^3}{2\hbar^2} \int [d\mathbf{k}] \sum_{mn\sigma} f_{mn} \Delta_{mn}^a \{r_{nm}^b, r_{mn}^c\} \delta(\omega_{nm} - \omega),$$

$$\eta_{\text{NIC}}^{abc} = -\frac{\pi e^3}{2\hbar^2} \int [d\mathbf{k}] \sum_{mn\sigma} f_{mn} \Delta_{mn}^a [r_{nm}^b, r_{mn}^c] \delta(\omega_{nm} - \omega).$$

$\eta_{\text{NIC}}^{abc}$ is the well-known NIC susceptibility, which vanishes under linearly polarized light. $\eta_{\text{MIC}}^{abc}$ is a new term, MIC susceptibility, arising from our microscopic derivation, which recovers the same formula as Yan's work [25]. It's clear that $\eta_{\text{MIC}}^{abc} = 0$ under time-reversal symmetry.

Similarly, we can separate shift current into the well-known NSC and a new term, MSC which has not been reported before:

$$J_{SC}^a = J_{NSC}^a + J_{MSC}^a = 2\sigma_{NSC}^{abc}\operatorname{Re}(E^b E^{c*}) + 2i\sigma_{MSC}^{abc}\operatorname{Im}(E^b E^{c*}),$$

where

$$\sigma_{NSC}^{abc} = -\frac{i\pi e^3}{4\hbar^2}\int[d\mathbf{k}]\sum_{mn\sigma} f_{nm}(r_{mn}^b r_{nm;k^a}^c + r_{mn}^c r_{nm;k^a}^b)(\delta(\omega_{nm}-\omega)+\delta(\omega_{mn}-\omega)),$$

$$\sigma_{MSC}^{abc} = -\frac{i\pi e^3}{4\hbar^2}\int[d\mathbf{k}]\sum_{mn\sigma} f_{nm}(r_{mn}^b r_{nm;k^a}^c - r_{mn}^c r_{nm;k^a}^b)(\delta(\omega_{nm}-\omega)+\delta(\omega_{mn}-\omega)).$$

Under time-reversal symmetry, the NSC and MSC susceptibilities can be rewritten in a compact form as follows,

$$\sigma_{NSC}^{abc} = -\frac{\pi e^3}{2\hbar^2}\int[d\mathbf{k}]\sum_{mn\sigma} f_{nm}R_{mn}^a(\mathbf{k})\{r_{nm}^b, r_{mn}^c\}\delta(\omega_{nm}-\omega),$$

$$\sigma_{MSC}^{abc} = -\frac{\pi e^3}{2i\hbar^2}\int[d\mathbf{k}]\sum_{mn\sigma} f_{nm}R_{mn}^a(\mathbf{k})\Omega_{mn}^a(\mathbf{k})\delta(\omega_{nm}-\omega),$$

where $R_{mn}^a(\mathbf{k}) = -\frac{\partial \phi_{mn}(\mathbf{k})}{\partial k^a} + \mathcal{A}_m^a(\mathbf{k}) - \mathcal{A}_n^a(\mathbf{k})$ is shift vector and $\phi_{nm}(\mathbf{k})$ is the phase factor of the interband Berry connection $r_{nm}^b(\mathbf{k}) = |r_{nm}^b(\mathbf{k})|e^{i\phi_{nm}(\mathbf{k})}$ and $\Omega_{mn}^a = i\epsilon_{abc}r_{mn}^b r_{nm}^c = i(r_{mn}^b r_{nm}^c - r_{mn}^c r_{nm}^b)$ is the local Berry curvature between band $m$ and $n$. The MSC is vanishing in time-reversal invariant system since Berry curvature $\Omega_{mn}^a$ is an odd function and shift vector is even function with $\mathbf{k}$ under time-reversal symmetry. MSC can be induced by circularly polarized light in gyrotropic magnetic materials. However, bilayer MnBi$_2$Te$_4$ of the following study is non-gyrotropic, hence it has vanishing MSC. Moreover, we can extend the above nonlinear shift and injection current to nonlinear shift spin current and nonlinear injection spin current by substituting the current operator $j^{a(s_b)} = -ev_a \to \frac{1}{2}\{s_b, v_a\}$, where $s_b = \frac{\hbar}{2}\sigma_b$ is spin operator with Pauli matrices $\sigma_b$. The extended nonlinear photo-spin current responses provide an essential route for exploring spin physics in condensed matter, which is currently under exploration and out of the scope in the present work.

**Crystal structure and group theoretical analysis of magnetic topological quantum material bilayer MnBi$_2$Te$_4$**

Here we present the magnetic group theoretical analysis of bilayer AFM-$z$ MnBi$_2$Te$_4$ (see Fig. 1a and 1c) with interlayer antiferromagnetic ordering of the Mn atoms aligned along the $z$ direction. In general, the dichromatic group $\mathcal{M}$ can be constructed by $\mathcal{M} = \mathcal{G}(\mathcal{H}) = \mathcal{H} \oplus (\mathcal{G} - \mathcal{H})\mathcal{T}$ where $\mathcal{G}$ is the ordinary geometric point group, $\mathcal{H}$ is the subgroup of $\mathcal{G}$ of index 2, and $\mathcal{T}$ is time-reversal operator. As a result, magnetic point group of bilayer AFM-$z$ MnBi$_2$Te$_4$ is $\bar{3}'m' = D_{3d}(D_3) = \{E, 2C_3, 3C_2', i\mathcal{T}, 2S_6\mathcal{T}, 3\sigma_d\mathcal{T}\}$. The character table of point group $D_{3d}$ is included in Table S1 in Supplementary Information. Without considering the ordering of spin, bilayer MnBi$_2$Te$_4$ is centrosymmetric and has no second-order nonlinear current responses. Taking into account the AFM-$z$ structure, $\mathcal{PT}$ symmetry is preserved in bilayer MnBi$_2$Te$_4$ with individual $\mathcal{P}$ and $\mathcal{T}$ symmetry being broken. In addition, bilayer AFM-$z$ MnBi$_2$Te$_4$ holds a two-fold rotation symmetry along $x$-axis $C_{2x} = \mathcal{M}_y\mathcal{M}_z$, and a combination of mirror symmetry and time-reversal symmetry $\mathcal{M}_x\mathcal{T}$, as shown in Fig. 1b. The electronic bands shown in Fig. 1d are doubly degenerate, which is generally true for $\mathcal{PT}$-symmetric system since $\mathcal{PT}E_{n,\uparrow}(\mathbf{k}) = E_{n,\downarrow}(\mathbf{k}) = E_{n,\uparrow}(\mathbf{k})$. Moreover, Figure 1e presents a contour plot of energy difference between the bottom conduction band and top valence band, which clearly shows the presence of three-fold rotation symmetry. As will be described below, all these symmetries play an important role in the nonlinear photocurrent responses. Bilayer MnBi$_2$Te$_4$ holds rich magnetic ordering, and the corresponding magnetic point group and symmetry operation of AFM-*x*, FM-*x*, and AFM-*z* structures are shown in Table 1. More details of group theoretical analysis of different magnetic structures can be found in Supplementary Information.

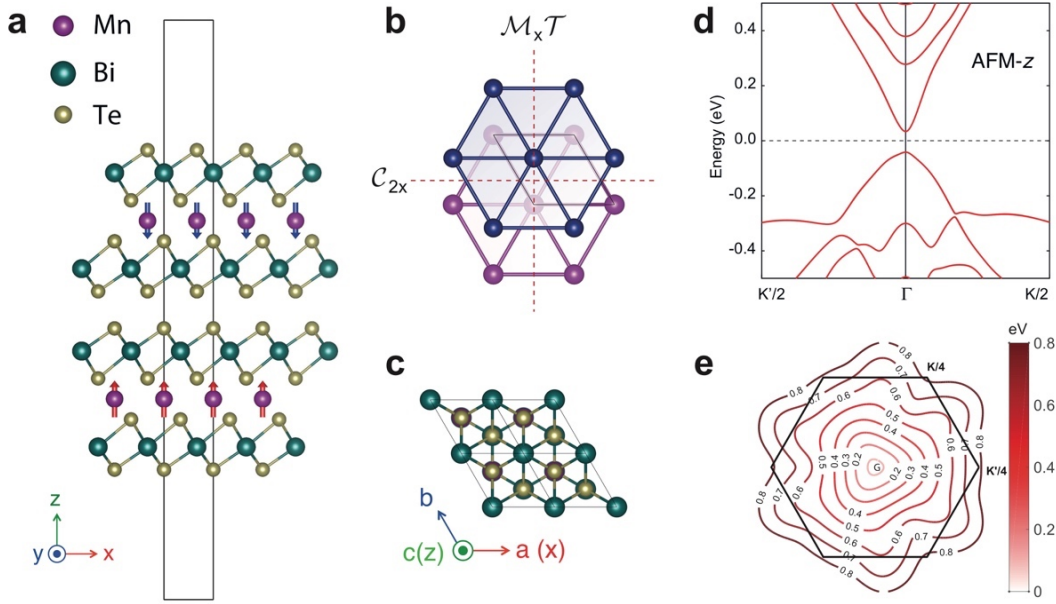

**Fig 1. | Crystal structure and electronic structure of bilayer AFM-*z* MnBi$_2$Te$_4$. a**, **c**. The side and top view of bilayer AFM-*z* MnBi$_2$Te$_4$. **b**. Selected magnetic symmetry elements in bilayer AFM-*z* MnBi$_2$Te$_4$ with the corresponding point group of $\bar{3}'m'$ in international notation or $D_{3d}(D_3)$ in Schoenflies notation. Two magnetic symmetry elements are illustrated in the plot, including two-fold rotation symmetry around *x*-axis $C_{2x} = \mathcal{M}_y\mathcal{M}_z$, and the combination of mirror symmetry and time-reversal symmetry $\mathcal{M}_x\mathcal{T}$. **d**, Electronic band structure of bilayer AFM-*z* MnBi$_2$Te$_4$ near the Fermi level. **e**, Contour plot of the energy difference between bottom conduction band and top valence band.

| Magnetic ordering | Symmetry operations | | | Magnetic point group |
|---|---|---|---|---|
| AFM-*z* | $\mathcal{PT}$ | $\mathcal{M}_x\mathcal{T}$ | $C_{2x} = \mathcal{M}_y\mathcal{M}_z$ | $\bar{3}'m'$ or $D_{3d}(D_3)$ |
| AFM-*x* | $\mathcal{PT}$ | $\mathcal{M}_x$ | $C_{2x}\mathcal{T} = \mathcal{M}_y\mathcal{M}_z\mathcal{T}$ | $2'/m$ or $C_{2h}(C_s)$ |
| FM-*z* | $\mathcal{P}$ | $\mathcal{M}_x\mathcal{T}$ | $C_{2x}\mathcal{T}$ | $\bar{3}m'$ or $D_{3d}(C_{3i})$ |
| FM-*x* | $\mathcal{P}$ | $\mathcal{M}_x$ | $C_{2x}$ | $2/m$ or $C_{2h}$ |

Table 1. Selected symmetry operations and magnetic point group in bilayer MnBi$_2$Te$_4$ with different magnetic ordering

For systems holding magnetic point group $\bar{3}'m'$ (such as bilayer and bulk AFM-$z$ MnBi$_2$Te$_4$), both normal magnetoelectric coupling and topological axion coupling are symmetry allowed. The polar $i$-vectors, e.g., electric field $\boldsymbol{E}$ or electric polarization $\boldsymbol{P}$, are represented by $A_{2u}(z)$ and $E_u(x,y)$ in $D_{3d}$ group and they are time reversal symmetric but space inversion antisymmetric. $A_{2g}(R_z)$ and $E_g(R_x, R_y)$ represent axial $i$-vectors which are symmetric under both time reversal and space inversion. The general transformation of polar/axial $i$-/$c$-tensors under space inversion $\mathcal{P}$, time reversal operation $\mathcal{T}$, and $\mathcal{PT}$ can be found in Table S2 in Supplementary Information. We can see the two types of vectors are incompatible in $D_{3d}$ group, e.g., $A_{2u} \otimes A_{2g} = A_{1u}$ does not contain total symmetric $A_{1g}$. The magnetic field $\boldsymbol{B}$ or orbital magnetization $\boldsymbol{M}$ are axial $c$-vectors which are time reversal antisymmetric but space inversion symmetric. For magnetic point groups, we use Birss's notation [26], where $\Gamma_s$ is the total symmetric representation, $\Gamma_m$ is the one-dimensional representation that corresponds to $\mathcal{M}$ in which all elements of $\mathcal{H}$ are represented by +1, and $\Gamma_p$ is the pseudovector representation in group $\mathcal{G}$. We shall write the direct product representation $\Gamma_{m \otimes p} = \Gamma_m \otimes \Gamma_p$ for axial $c$-vector. For magnetic point group $D_{3d}(D_3)$, we have $\Gamma_s = A_{1g}, \Gamma_m = A_{1u}, \Gamma_p = A_{2g} + E_g$. Thus, $\Gamma_{m \otimes p} = A_{2u} + E_u$. Consequently, $\Gamma_M \otimes \Gamma_E = \Gamma_P \otimes \Gamma_B = 2A_{1g} + A_{2g} + 3E_g$, which contains $2A_{1g}$. Hence, the linear magnetoelectric coupling of $\bar{3}'m'$ structures has two independent nonvanishing components and corresponding orbital magnetoelectric polarizability. Now let's see how $c$-type second-order current response under linearly polarized light such as MIC is enabled in $D_{3d}(D_3)$ by considering the following direct product $\Gamma_m \otimes \Gamma_j \otimes (\Gamma_E \otimes \Gamma_E)^S = A_{1g} + 3A_{2g} + 4E_g$, indicating only one independent and nonvanishing susceptibility is allowed. For $i$-type second-order current response under linearly polarized light such as NSC, we have the direct product: $\Gamma_j \otimes (\Gamma_E \otimes \Gamma_E)^S = (A_{2u} + E_u) \otimes (2A_{1g} + E_g) = A_{1u} + 3A_{2u} + 4E_u$. Hence, NSC is not symmetry allowed. In fact, NSC is immune to the time reversal symmetry and can be determined by nonmagnetic point group. As bilayer/bulk MnBi$_2$Te$_4$ holds nonmagnetic point group $D_{3d}$ with inversion symmetry, both NSC and NIC vanish. Similarly, $c$-type second-order current response under circularly polarized light such as MSC is forbidden in $D_{3d}(D_3)$ by considering the following direct product: $\Gamma_m \otimes \Gamma_j \otimes \Gamma_{E \times E^*} = 2A_{1u} + A_{2u} + 3E_u$ which does not contain total symmetric representation. The above magnetic group theoretical analysis can be extended to other higher order optical/photocurrent responses in magnetic materials.

**Magnetic injection current and normal shift current in magnetic topological quantum material bilayer MnBi$_2$Te$_4$**

Figure 2a and 2b present the electronic structure and MIC susceptibility $\eta_{\text{MIC}}^{xxx}$ in bilayer MnBi$_2$Te$_4$ with magnetic ordering in AFM-$z$ (↑↓) and time-reversed AFM-$z$ (tAFM-$z$, ↓↑), respectively. The corresponding magnetic ordering is shown in Fig. 2a inset. The $\boldsymbol{k}$-space distribution of $f_{mn}\Delta_{mn}^a\{r_{nm}^b, r_{mn}^c\}\delta(\omega_{nm} - \omega)$ with a unit of Å$^3$ at $\omega = 120$ meV and $\omega = 600$ meV are shown in Figure 2c and 2d for bilayer AFM-$z$ and in Fig. 2e and 2f for tAFM-$z$ MnBi$_2$Te$_4$, respectively. This term arises from the large difference of the asymmetric group velocity $\Delta_{mn}^a$ and the absorption strength $\{r_{nm}^b, r_{mn}^c\}$ at time-reversed $\pm\boldsymbol{k}$ points, which eventually contributes to the large MIC in bilayer AFM-$z$ and tAFM-$z$ MnBi$_2$Te$_4$. The peak MIC conductivity is $\sim 2 \times 10^{10}$ nm · A/(V$^2$ · s), two times the peak NIC in ferroelectric monolayer GeS [17]. More importantly, it demonstrates that the MIC can be switched upon magnetic ordering transition between AFM-$z$ (↑↓) and its time-reversed tAFM-$z$ (tAFM, ↓↑) configurations. With additional three-fold rotation symmetry, we have $\eta_{\text{MIC}}^{xxx} = -\eta_{\text{MIC}}^{xyy} = -\eta_{\text{MIC}}^{yxy} = -\eta_{\text{MIC}}^{yyx}$. Moreover, the presence of $\mathcal{M}_y\mathcal{M}_z$ symmetry leads to vanishing $\eta_{\text{MIC}}^{yyy}$.

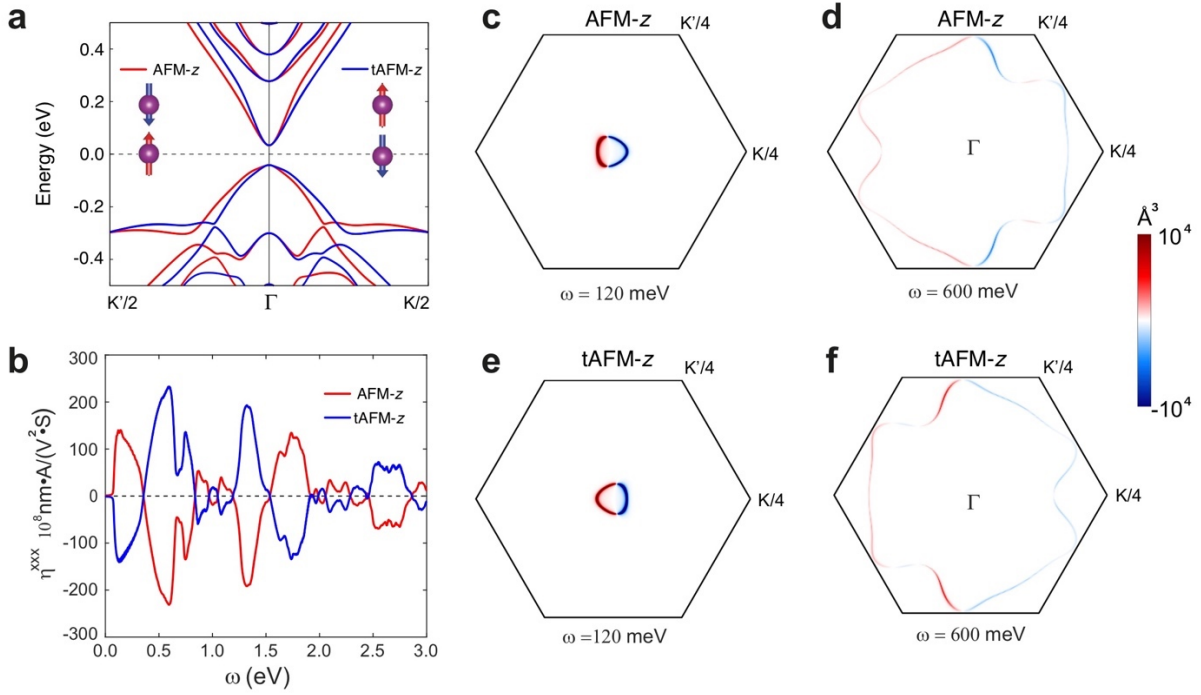

**Fig 2. | Magnetic injection current (MIC) in bilayer MnBi$_2$Te$_4$ with magnetic ordering in AFM-$z$ (↑↓) and time-reversed AFM-$z$ (tAFM, ↓↑). a**, Band structure of bilayer AFM-$z$ and tAFM-$z$ MnBi$_2$Te$_4$ with spin-orbit coupling taken into account. **b**, Frequency-dependent magnetic injection current response of bilayer AFM-$z$ and tAFM-$z$ MnBi$_2$Te$_4$. **c-f**, Microscopic distribution of MIC $f_{mn}\Delta_{mn}^a\{r_{nm}^b, r_{mn}^c\}\delta(\omega_{nm} - \omega)$ between top valence band and bottom conduction band at different frequencies of 120 meV and 600 meV in bilayer AFM-$z$ and tAFM-$z$ MnBi$_2$Te$_4$. It shows that MIC can be switched between two time-reversed magnetic orders AFM-$z$ (↑↓) and tAFM (↓↑).

Bilayer MnBi$_2$Te$_4$ may have magnetic ordering different from the above AFM-$z$ structure, consequently the corresponding nonlinear current responses such as NSC and MIC can be very different. For example, FM-$x$ and FM-$z$ magnetic ordering under external magnetic field (*i.e.* ferromagnetic ordering with magnetization aligned along $x/z$ direction) holds inversion symmetry, hence has vanishing even-order nonlinear optical and photocurrent responses including SHG, NSC, NIC, MSC, and MIC. For example, Figure S1 shows the band structure for the FM-$z$ magnetic ordering where each band is symmetric with respect to ±$\boldsymbol{k}$ points with vanishing MIC. In addition, for bilayer MnBi$_2$Te$_4$ with AFM-$x$ magnetic ordering, the corresponding magnetic point group becomes $2'/m = C_{2h}(C_s) = C_s \oplus (C_{2h} - C_s)\mathcal{T} = \{E, \sigma_x, C_{2x}\mathcal{T}, i\mathcal{T}\}$, different from $\bar{3}'m'$ for bilayer MnBi$_2$Te$_4$ with AFM-$z$ magnetic structure. Nonetheless, bilayer AFM-$x$ MnBi$_2$Te$_4$ preserves $\mathcal{PT}$-symmetry and allows for nonvanishing MIC response $\eta_{\text{MIC}}^{xxy}$, $\eta_{\text{MIC}}^{yxx}$ and $\eta_{\text{MIC}}^{yyy}$ as shown in Fig. S2 in Supplementary Information. It is worth to point out that van der Waals layered MnBi$_2$Te$_4$ hosts rich topology with varying magnetic ordering. Probing MIC response can, therefore, help directly determine the intrinsic magnetic structure and hence potential topology of magnetic topological quantum materials owing to the intimate coupling between magnetic symmetry and nonlinear photocurrent responses.

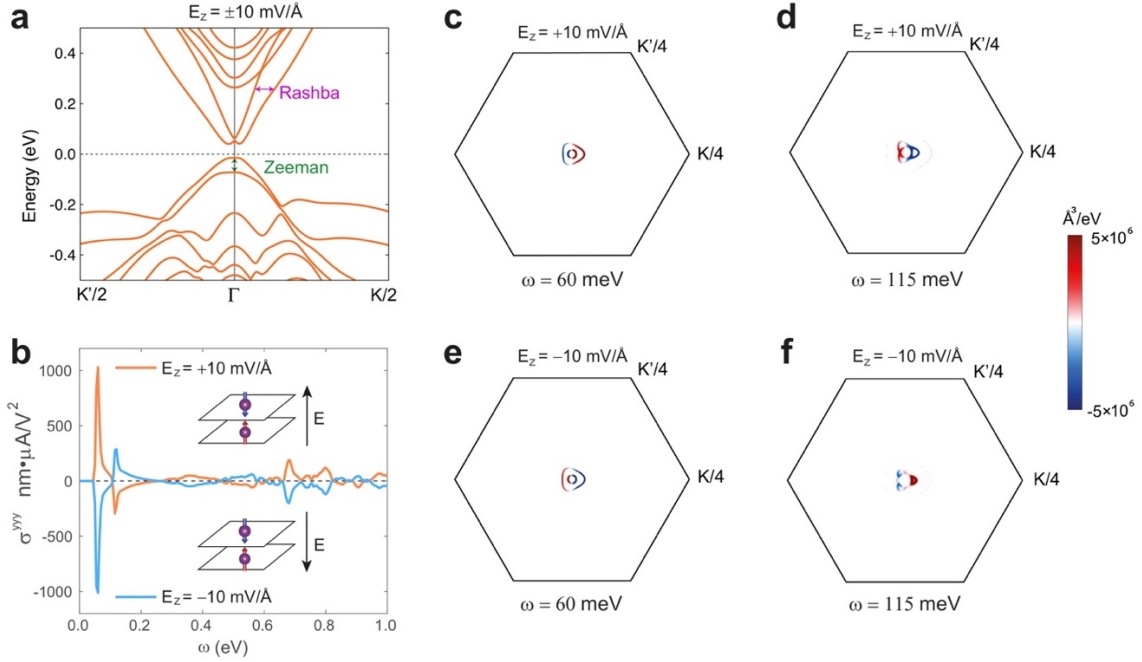

**Fig. 3. | Normal shift current (NSC) in bilayer MnBi$_2$Te$_4$ with magnetic ordering AFM-*z* (↑↓) under external electric field $E_z = 10$ mV/Å. a**, Band structure of bilayer AFM-*z* under external electric field $E_z = 10$ mV/Å. **b**, Frequency-dependent NSC response of bilayer AFM-*z* MnBi$_2$Te$_4$ under external electric field $E_z = \pm 10$ mV/Å. **c-f**, Microscopic distribution of NSC $\sum_{mn\sigma} f_{nm}(r^b_{mn} r^c_{nm;k^a} + r^c_{mn} r^b_{nm;k^a})(\delta(\omega_{nm} - \omega) + \delta(\omega_{mn} - \omega))$ at different frequencies in bilayer AFM-*z* external electric field $E_z = \pm 10$ mV/Å. It demonstrates that NSC can be switched by applying electric field along +z/-z out-of-plane directions.

Although both bilayer AFM-*z* and tAFM-*z* MnBi$_2$Te$_4$ hold the large MIC response as shown above, their Berry curvature $\Omega_n(\mathbf{k})$ and shift vector $R^a_{mn}(\mathbf{k})$ are vanishing due to $\mathcal{PT}\Omega_n(\mathbf{k}) = -\Omega_n(\mathbf{k}) = \Omega_n(\mathbf{k})$ and $\mathcal{PT} R^a_{mn}(\mathbf{k}) = -R^a_{mn}(\mathbf{k}) = R^a_{mn}(\mathbf{k})$. Hence, no NSC and Berry curvature related physics would be expected in $\mathcal{PT}$-symmetric system. In fact, one can take the wave functions to be real (due to the $\mathcal{PT}$ transformation). Then the phase effect is absent and the shift vector and Berry curvature become zero in $\mathcal{PT}$-symmetric system. However, one can apply Berry curvature engineering via electric gating to break the $\mathcal{PT}$ symmetry symmetry in the pristine bilayer MnBi$_2$Te$_4$ [28]. Here we apply an external electric field along the out of plane direction to bilayer MnBi$_2$Te$_4$. Figure 3a and 3b show the calculated electronic structure and NSC in bilayer MnBi$_2$Te$_4$ under an external electric field $E_z = 10$ mV/Å with magnetic ordering AFM-z (↑↓). Large Rashba and Zeeman spin splitting are indicated in Figure 3a which we will discuss soon. The peak NSC conductivity is $\sim 1000$ nm·$\mu$A/V$^2$, which is 50 times larger than that in GeS [17]. In addition, Figure 3c-3f present the microscopic distribution of NSC, $\sum_{mn\sigma} f_{nm}(r^b_{mn} r^c_{nm;k^a} + r^c_{mn} r^b_{nm;k^a})(\delta(\omega_{nm} - \omega) + \delta(\omega_{mn} - \omega))$, with a unit of Å$^3$/eV at two frequencies ($\omega = 60$ meV and $\omega = 115$ meV) in bilayer AFM-$z$ with $E_z = \pm 10$ mV/Å. From the symmetry transformation $\mathcal{M}_y \mathcal{M}_z E_z = -E_z$, we can see that the direction of NSC along $y$ under linearly $y$-polarized light can be flipped by switching the electric field from +$z$ to -$z$.

Similar to the case of MIC, $\sigma^{yyy}_{NSC} = -\sigma^{yxx}_{NSC} = -\sigma^{xyx}_{NSC} = -\sigma^{xxy}_{NSC}$ under three-fold rotation symmetry, and $\sigma^{xxx}_{NSC}$ vanishes due to $\mathcal{M}_x \mathcal{T}$ symmetry. It's important to realize that upon linearly polarized incident light along $x$ direction, the response of NSC in the electrically gated bilayer AFM-$z$ and tAFM-$z$ MnBi$_2$Te$_4$ is governed by $\sigma^{yxx}_{NSC}$ with nonlinear photocurrent generated along $y$ only. In contrast, under the same linearly $x$-polarized light, the response of MIC is dominated by $\eta^{xxx}_{MIC}$ with nonlinear photocurrent generated along $x$ only. Moreover, NSC does not switch in bilayer AFM-$z$ (↑↓) and time-reversed AFM-$z$ (tAFM, ↓↑) MnBi$_2$Te$_4$, indicating that NSC cannot be reversed under $\mathcal{T}$ corresponding to its dissipative and non-reciprocal nature [29]. In addition, as aforementioned, NSC vanishes when electric field is absent, while MIC always exists regardless of electric field. As shown in Figure 2 and Figure 3, both NSC and MIC in bilayer AFM-z MnBi$_2$Te$_4$ have strong responses at low energy regime (e.g. two strong NSC peaks at 60 meV and 115 meV under $E_z = 10$ mV/Å and two strong MIC peaks at 120 meV and 600 meV), making them particularly attractive for terahertz and infrared sensing.

Both NSC and MIC are highly tunable under varying electric field. Figure 4a shows the electric field dependent band structure. The band gap decreases upon increasing vertical electric field until a critical field of $E_z = 25$ mV/Å where it undergoes topological phase transition from zero-plateau quantum anomalous Hall insulator with Chern number $C_n = 0$ to high Chern number quantum anomalous Hall insulator ($C_n = 3$). The nontrivial topology is evinced in the Berry curvature shown in Fig. 4b which is localized around the center of the Brillouin zone. The integration of Berry curvature yields a high Chern number of $C_n = 3$. As aforementioned, electric field can introduce large Rashba and Zeeman splitting in bilayer AFM-$z$ MnBi$_2$Te$_4$. In Fig. 4c and 4d we show the corresponding spin polarization at the energy of 0.6 eV with AFM-$z$ and tAFM-$z$ magnetic ordering under a vertical electric field of $E_z = 10$ mV/Å, which confirms large spin splitting with three-fold rotation symmetry under finite electric field. Under the same electric field, spin polarization in AFM-$z$ and tAFM-$z$ are related by time reversal $\mathcal{T}$ operation. However, for the same magnetic configuration AFM-$z$ or tAFM-$z$, the spin polarization under electric field along +z and -z are related by $\mathcal{PT}$ operation (see Fig. S3 in Supplementary Information). Electric field induced large spin splitting and distinct spin polarization suggest that bilayer AFM-$z$ MnBi$_2$Te$_4$ could serve as a rich platform for developing electrically controlled spintronics.

Electric field induced Rashba and Zeeman splitting reduces the electronic gap, thereby shifting the nonlinear photocurrent responses such as MIC and NSC to the THz regime. Figure 4e and 4f show field-dependent NSC and MIC responses in bilayer AFM-$z$ MnBi$_2$Te$_4$ under varying electric field. It clearly demonstrates that the strength of NSC response increases under increasing electric field until $E_z = 23$ mV/Å, and the MIC conductibility remains very large until $E_z = 23$ mV/Å. More importantly, the low-energy peak of NSC and MIC response linearly decreases to ~25 meV upon increasing field up to $E_z =$ 23 mV/Å. It's worth to point out that NSC will switch the sign upon the switching of electric field from +z to -z, however the sign of MIC will remain the same. This is another way to distinguish NSC and MIC. We would like to mention that the NSC and MIC calculated here is solely contributed by bulk response. Topologically protected edge states in QAHI will also contribute nonlinear current responses, which is outside the scope of this work and needs to be explored in future.

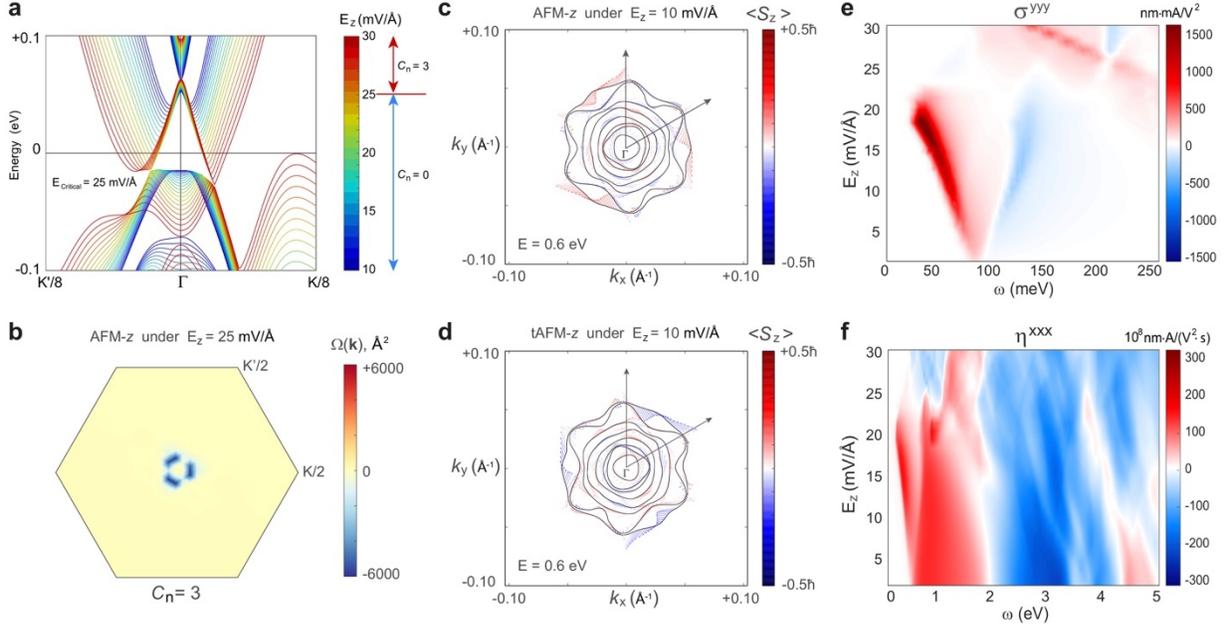

**Fig. 4. | Electric field dependent electric structure, topology, and photocurrent responses in bilayer AFM-$z$ MnBi$_2$Te$_4$.** **a**, Electric field dependent band structure. **b**, Berry curvature at the critical vertical electric field of $E_z = 25$ mV/Å where it undergoes topological phase transition from zero-plateau quantum anomalous Hall insulator with Chern number $C_n = 0$ to high Chern number quantum anomalous Hall insulator ($C_n = 3$). **c**, **d**, Spin polarization of bilayer MnBi$_2$Te$_4$ at energy of 0.6 eV with AFM-$z$ and tAFM-$z$ magnetic ordering under a vertical electric field of $E_z = 10$ mV/Å which enables large spin splitting. **e**, Normal shift photocurrent (NSC) susceptibility, $\sigma_{\text{NSC}}^{yyy}$, in bilayer AFM-$z$ MnBi$_2$Te$_4$ under varying external static vertical electric field $E_z$ and fundamental frequency $\omega$. **f**, Magnetic injection current (MIC) susceptibility, $\eta_{\text{MIC}}^{xxx}$, in bilayer AFM-$z$ MnBi$_2$Te$_4$ under varying external static vertical electric field $E_z$ and fundamental frequency $\omega$.

## Discussion

The distinct directional dependence and switching behavior of NSC and MIC under different electric field and magnetic ordering provide a fundamental basis to distinguish these two current responses in experiment. First, as shown above, under linearly polarized light, MIC can be generated in $\mathcal{PT}$-symmetric system such as bilayer AFM-*z* MnBi$_2$Te$_4$. NSC, though absent in the $\mathcal{PT}$-symmetric system, can be enabled via electric gating. Moreover, for electrically gated bilayer AFM-*z* and tAFM-*z* MnBi$_2$Te$_4$ under linearly *x*-polarized light, NSC susceptibility tensor element $\sigma_{\text{NSC}}^{yxx}$ contributes a giant nonlinear photocurrent along *y* only, while MIC susceptibility tensor element $\eta_{\text{MIC}}^{xxx}$ contributes a large nonlinear photocurrent along *x* only. More importantly, nonlinear photocurrent from NSC can be switched by applying vertical electric field along +*z*/-*z* direction, while nonlinear photocurrent from MIC can be reversed by inducing magnetic transition between AFM-*z* and tAFM-*z*. Furthermore, different magnetic orderings (e.g. AFM-*z*, AFM-*x*, FM-*z*, and FM-*x*) yield distinct magnetic point group, thus different directional dependence of nonlinear photocurrent. Such rich coupling between the magnetic ordering and nonlinear photocurrent responses presents itself a powerful tool for investigating magnetic symmetries, structures, and interactions in magnetically ordered crystals as well as imaging magnetic domain walls.

Bilayer AFM-*z* MnBi$_2$Te$_4$ possesses giant MIC and NSC that are highly tunable under electric field with the first low-energy peak red-shifted down to 25 meV (*i.e.* ~6 THz), offering a promising platform for THz sensing of extreme nonlinear quantum phenomena. In addition to MIC and NSC, MSC presents another interesting second order photocurrent response which vanishes in time-reversal invariant systems. MSC can be induced by circularly polarized light in gyrotropic magnetic materials. Bilayer MnBi$_2$Te$_4$ in the present study is however non-gyrotropic, hence MSC vanishes. MSC in gyrotropic magnetic materials is worth for a further exploration. The present work did not consider the third-order current response under static electric field. During the preparation of this manuscript, we noticed that another paper [30] by Fei *et al*. also proposed giant linearly polarized photogalvanic effect in MnBi$_2$Te$_4$ using a different approach which is equivalent to MIC. Nonlinear shift and injection current can be extended to nonlinear shift spin current and nonlinear injection spin current, yielding nonlinear photo-spin current response even in centrosymmetric systems. This will provide an essential route for exploring spin physics in condensed matter, which is currently under exploration and out of the scope in the present work.

**Conclusion**

The intimate coupling between intrinsic magnetic point group and nonlinear optical/photocurrent responses makes nonlinear spectroscopy and imaging a powerful tool for investigating magnetic symmetries, structures and interactions in magnetically ordered crystals. In particular, we proposed two new types of second-order nonlinear direct photocurrents, MSC and MIC, as the counterparts of NSC and NIC in time-reversal symmetry and inversion symmetry broken system. Taking magnetic topological quantum material in MnBi$_2$Te$_4$ as an example, we predicted that $\mathcal{PT}$-symmetric bilayer MnBi$_2$Te$_4$ possesses giant, magnetically switchable MIC in the THz regime. Although NSC vanishes in $\mathcal{PT}$ symmetric system, electric field breaks the $\mathcal{PT}$-symmetry in bilayer AFM-*z* MnBi$_2$Te$_4$, and enables large NSC response at the IR regime. More importantly, NSC can be switched by vertical electric field and MIC be switched upon magnetic transition, however NSC (MIC) will not be switched by magnetic transition (electric field). Due to the magnetic group symmetry of bilayer AFM-*z* MnBi$_2$Te$_4$, MIC and NSC yield photocurrent that are perpendicular to each other upon linearly *x/y*-polarized light, hence can be distinguished. Very excitingly, electric gating can efficiently tune the large nonlinear photocurrent response down to 6 THz, suggesting bilayer AFM-*z* MnBi$_2$Te$_4$ as an exciting platform for THz and magneto-optoelectronic applications as well as 2D spintronics that are electrically and magnetically tunable. The present work reveals that nonlinear photocurrent responses governed by NSC, NIC, MSC, and MIC provide a powerful spectroscopic/imaging tool for the investigation of magnetic structures and interactions which could be particularly fruitful for probing and understanding magnetic topological quantum materials.

## Methods

**Ground state crystal and electronic structure.** Ground-state crystal structures of $MnBi_2Te_4$ were calculated using first-principles density-functional theory [31,32] implemented in the Vienna Ab initio Simulation Package (VASP) [33,34] with the projector-augmented wave method [35] and a plane-wave basis with an energy cutoff of 400 eV. We employed the generalized-gradient approximation of exchange-correlation energy functional in the Perdew-Burke-Ernzerhof [36] form. A Hubbard U correction with U= 4 eV was applied to Mn atoms to reduce the self-interaction error present in DFT-PBE. We further adopted a Monkhorst-Pack k-point sampling of 11×11×1 for the Brillouin zone integration and DFT-D3 functional [37] to account for weak interlayer dispersion interactions. The convergence criteria for maximal residual force was set to 0.005 eV/Å, and the convergence criteria for electronic relaxation was set to $10^{-6}$ eV.

**First-principles calculations of nonlinear photocurrent.** With the fully relaxed ground-state crystal structure, we constructed quasiatomic spinor Wannier functions and tight-binding Hamiltonian from Kohn-Sham wavefunctions and eigenvalues under the maximal similarity measure with respect to pseudoatomic orbitals [38,39]. Spin-orbit coupling was taken into account. Total 120 quasiatomic spinor Wannier functions were obtained from the projections onto Mn's *s* and *d* pseudo-atomic orbitals, Te's *s* and *p* pseudo-atomic orbitals, and Bi's *s* and *p* pseudo-atomic orbitals for bilayer $MnBi_2Te_4$. Using the developed tight-binding Hamiltonian, we then computed MIC and NSC susceptibility tensor using a modified WANNIER90 code [40]. A small imaginary smearing factor $\eta$ of 5 meV was applied to fundamental frequency. A k-point grid of 1000×1000×1 was adopted, which is dense enough to the convergence as shown in Fig. S4.

## Data availability
The data that support the plots within this paper and other findings of this study are available from the corresponding author upon reasonable request.

## References
1. Mong, R. S., Essin, A. M. & Moore, J. E. Antiferromagnetic topological insulators. *Physical Review B* **81**, 245209 (2010).
2. Qi, X.-L., Hughes, T. L. & Zhang, S.-C. Topological field theory of time-reversal invariant insulators. *Physical Review B* **78**, 195424 (2008).
3. Zhang, D. *et al.* Topological Axion States in the Magnetic Insulator MbBi2Te4 with the Quantized Magnetoelectric Effect. *Phys. Rev. Lett.* **122**, 206401 (2019).
4. Li, J. *et al.* Intrinsic magnetic topological insulators in van der Waals layered MnBi2Te4-family materials. *Science Advances* **5**, eaaw5685 (2019).
5. Otrokov, M. M. *et al.* Unique Thickness-Dependent Properties of the van der Waals Interlayer Antiferromagnet MnBi2Te4 Films. *Phys. Rev. Lett.* **122**, 107202 (2019).

6  Otrokov, M. M. *et al.* Prediction and observation of an antiferromagnetic topological insulator. *Nature* **576**, 416-422 (2019).
7  Liu, C. *et al.* Robust axion insulator and Chern insulator phases in a two-dimensional antiferromagnetic topological insulator. *Nat. Mater.* **19**, 522-527 (2020).
8  Gong, Y. *et al.* Experimental realization of an intrinsic magnetic topological insulator. *Chinese Physics Letters* **36**, 076801 (2019).
9  Thomson, R., Leburn, C. & Reid, D. *Ultrafast Nonlinear Optics*. (Springer, 2013).
10 Kwiat, P. G., Waks, E., White, A. G., Appelbaum, I. & Eberhard, P. H. Ultrabright source of polarization-entangled photons. *Phys. Rev. A* **60**, R773-R776 (1999).
11 Gisin, N., Ribordy, G., Tittel, W. & Zbinden, H. Quantum cryptography. *Rev. Mod. Phys.* **74**, 145-195 (2002).
12 Zhao, M. *et al.* Atomically phase-matched second-harmonic generation in a 2D crystal. *Light: Science and Applications* **5**, e16131 (2016).
13 Wang, H. & Qian, X. Giant Optical Second Harmonic Generation in Two-Dimensional Multiferroics. *Nano Lett.* **17**, 5027-5034 (2017).
14 Sipe, J. & Shkrebtii, A. Second-order optical response in semiconductors. *Physical Review B* **61**, 5337 (2000).
15 Young, S. M. & Rappe, A. M. First Principles Calculation of the Shift Current Photovoltaic Effect in Ferroelectrics. *Phys. Rev. Lett.* **109**, 116601 (2012).
16 Rangel, T. *et al.* Large Bulk Photovoltaic Effect and Spontaneous Polarization of Single-Layer Monochalcogenides. *Phys. Rev. Lett.* **119**, 067402 (2017).
17 Wang, H. & Qian, X. Ferroicity-driven nonlinear photocurrent switching in time-reversal invariant ferroic materials. *Science Advances* **5**, eaav9743 (2019).
18 Nastos, F. & Sipe, J. Optical rectification and current injection in unbiased semiconductors. *Physical Review B* **82**, 235204 (2010).
19 Panday, S. R., Barraza-Lopez, S., Rangel, T. & Fregoso, B. M. Injection current in ferroelectric group-IV monochalcogenide monolayers. *Physical Review B* **100**, 195305 (2019).
20 Sodemann, I. & Fu, L. Quantum Nonlinear Hall Effect Induced by Berry Curvature Dipole in Time-Reversal Invariant Materials. *Phys. Rev. Lett.* **115**, 216806 (2015).
21 Moore, J. E. & Orenstein, J. Confinement-Induced Berry Phase and Helicity-Dependent Photocurrents. *Phys. Rev. Lett.* **105**, 026805 (2010).
22 Wang, H. & Qian, X. Ferroelectric nonlinear anomalous Hall effect in few-layer $WTe_2$. *npj Computational Materials* **5**, 119 (2019).
23 Xiao, J. *et al.* Berry curvature memory via electrically driven stacking transitions. *arXiv preprint arXiv:1912.01037* (2019).
24 Fregoso, B. M. Bulk photovoltaic effects in the presence of a static electric field. *Physical Review B* **100**, 064301 (2019).
25 Zhang, Y. *et al.* Switchable magnetic bulk photovoltaic effect in the two-dimensional magnet $CrI_3$. *Nat. Commun.* **10**, 3783 (2019).
26 Birss, R. R. *Symmetry and magnetism*. Vol. 863 (North-Holland Amsterdam, 1964).
27 Mahon, P. T. & Sipe, J. Magnetoelectric polarizability: A microscopic perspective. *arXiv preprint arXiv:2003.00313* (2020).
28 Du, S. *et al.* Berry curvature engineering by gating two-dimensional antiferromagnets. *Physical Review Research* **2**, 022025 (2020).
29 Morimoto, T. & Nagaosa, N. Nonreciprocal current from electron interactions in noncentrosymmetric crystals: roles of time reversal symmetry and dissipation. *Scientific Reports* **8**, 2973 (2018).
30 Fei, R., Song, W. & Yang, L. Giant linearly-polarized photogalvanic effect and second harmonic generation in two-dimensional axion insulators. *arXiv preprint arXiv:2003.01576* (2020).
31 Hohenberg, P. & Kohn, W. Inhomogeneous electron gas. *Phys. Rev. B* **136**, B864-B871 (1964).
32 Kohn, W. & Sham, L. J. Self-consistent equations including exchange and correlation effects. *Phys. Rev.* **140**, A1133-A1138 (1965).

33. Kresse, G. & Furthmüller, J. Efficient iterative schemes for *ab initio* total-energy calculations using a plane-wave basis set. *Physical Review B* **54**, 11169-11186 (1996).
34. Kresse, G. & Furthmüller, J. Efficiency of ab-initio total energy calculations for metals and semiconductors using a plane-wave basis set. *Comput. Mater. Sci.* **6**, 15-50 (1996).
35. Blöchl, P. E. Projector augmented-wave method. *Phys. Rev. B* **50**, 17953-17979 (1994).
36. Perdew, J. P., Burke, K. & Ernzerhof, M. Generalized gradient approximation made simple. *Phys. Rev. Lett.* **77**, 3865-3868 (1996).
37. Grimme, S., Antony, J., Ehrlich, S. & Krieg, H. A consistent and accurate ab initio parametrization of density functional dispersion correction (DFT-D) for the 94 elements H-Pu. *J. Chem. Phys*. **132**, 154104 (2010).
38. Marzari, N., Mostofi, A. A., Yates, J. R., Souza, I. & Vanderbilt, D. Maximally localized Wannier functions: Theory and applications. *Rev. Mod. Phys.* **84**, 1419-1475 (2012).
39. Qian, X. *et al.* Quasiatomic orbitals for *ab initio* tight-binding analysis. *Phys. Rev. B* **78**, 245112 (2008).
40. Mostofi, A. A. *et al.* An updated version of wannier90: A tool for obtaining maximally-localised Wannier functions. *Comput. Phys. Commun.* **185**, 2309-2310 (2014).

**Acknowledgments**

We thank Yong Xu, Shiqiao Du, and Takahiro Morimoto for helpful discussions. This work was supported by the National Science Foundation (NSF) under award number DMR-1753054. Portions of this research were conducted with the advanced computing resources provided by Texas A&M High Performance Research Computing.

**Author contributions**

X.Q. supervised the project; H.W. and X.Q. conceived the research. H.W. developed first-principles code for magnetic injection current susceptibility and carried out the calculations. H.W. and X.Q. conducted theoretical analysis, analyzed the results, and wrote the manuscript.

**Competing interests**

The authors declare no competing interests.